\documentclass[12pt,aps,prd,preprint,tightenlines,superscriptaddress,
   showpacs,nofootinbib]{revtex4-1}
\usepackage{graphicx}

\usepackage{amsmath}
\usepackage{array}

\newcommand{\PRE}[1]{{#1}} 

\newcommand{\nbox}{{\,\lower0.9pt\vbox{\hrule \hbox{\vrule height 0.2 cm
\hskip 0.2 cm \vrule height 0.2 cm}\hrule}\,}}

\newcommand{\gev}{\text{GeV}}

\newcommand{\be}{\begin{equation}}
\newcommand{\ee}{\end{equation}}
\newcommand{\bea}{\begin{eqnarray}}
\newcommand{\eea}{\end{eqnarray}}
\newcommand{\baln}{\begin{align}}
\newcommand{\ealn}{\end{align}}
\newcommand{\lsim}{\lower.7ex\hbox{$\;\stackrel{\textstyle<}{\sim}\;$}}
\newcommand{\gsim}{\lower.7ex\hbox{$\;\stackrel{\textstyle>}{\sim}\;$}}

\usepackage{graphicx}
\begin{document}

\preprint{UH-511-1208-2013}
\preprint{CETUP*12-020}

\title{
\PRE{\vspace*{1.3in}}
Gamma Rays from Bino-like Dark Matter in the MSSM
\PRE{\vspace*{0.3in}}
}

\author{Jason Kumar}
\affiliation{Department of Physics and Astronomy, University of
Hawai'i, Honolulu, HI 96822, USA
\PRE{\vspace*{.1in}}
}

\author{Pearl Sandick%
\PRE{\vspace*{.4in}}
}
\affiliation{Department of Physics and Astronomy, University of Utah, Salt Lake City, UT  84112, USA
\PRE{\vspace*{.5in}}
}


\begin{abstract}
\PRE{\vspace*{.3in}}
We consider regions of the parameter space of the Minimal Supersymmetric
Standard Model (MSSM) with bino-like neutralino dark matter in which a large fraction of the total dark matter
annihilation cross section in the present era arises from annihilation to final states with monoenergetic photons.
The region of interest
is characterized by light sleptons and heavy squarks.
We find that it is possible for the branching fraction to final states with monoenergetic photons to
be comparable to that for continuum photons, but in those cases the total cross section will be so small that
it is unlikely to be observable.  For models where dark matter annihilation
may be observable in the present era, the branching fraction to final states with monoenergetic photons is ${\cal O}(1-10\%)$.
\end{abstract}


\maketitle

\section{Introduction} \label{sec:intro}

In light of recent analyses of data from the Fermi Large Area Telescope (LAT)~\cite{FermiLAT} indicating the
possibility of an excess of $\sim 130~\gev$ photons from the
Galactic Center region~\cite{FermiLine}, there has been renewed interest in dark matter models
that can produce monoenergetic photons through the processes $\chi \chi \rightarrow
\gamma \gamma, \gamma Z,$ and $\gamma h$~\cite{buckleyhooper, FermiModels, shakya, FermiModelsMSSM, Cohen:2012me, FermiModelsNMSSM, OtherModels}.
Generically, one would expect models with a monoenergetic photon line to be
accompanied by a larger continuum photon spectrum.  This reasoning derives from
general principles: since there are tight constraints on the electromagnetic
charge of dark matter, monoenergetic photon production can only arise through
a millicharged coupling or through a loop diagram.  These processes are expected
to be suppressed relative to tree-level annihilation to other Standard Model particles.
The dominant annihilation products would then produce a continuum photon flux,
either from decays that produce photons or through synchrotron radiation, which is large
relative to the strength of a monoenergetic photon signal.

Bounds on the magnitude of the continuum spectrum arising from dark matter annihilation
in the Galactic Center thus put tight constraints on dark matter models that could
explain a line signal.  Many new dark matter models have been proposed to evade these constraints
and explain the recent Fermi data~\cite{buckleyhooper, FermiModels}.  Similarly, a variety of works have studied the question
of whether it is possible for
the $130~\gev$ photon excess seen by
Fermi to be produced within the MSSM~\cite{shakya, FermiModelsMSSM, Cohen:2012me} or singlet extensions of the MSSM~\cite{FermiModelsNMSSM},
while maintaining consistency with bounds on the continuum spectrum.

In this work, we consider an MSSM analysis from a different perspective.  We assume that the MSSM lightest supersymmetric particle (LSP)
is the lightest neutralino, $\chi$, and constitutes all
of the dark matter in the universe, and we ask what we can learn about
the parameters of the MSSM from the observation of a photon line signal and a small (or unobservable)
continuum signal.  With this goal in mind, we avoid (as far as possible) any assumptions about
the MSSM parameter space based on naturalness or simplicity, instead taking as our guide only the
mass of the Higgs boson (taken to be $m_h \sim 126~\gev$), other observational constraints, and the assumption
of a large photon line signal and small continuum signal.  We will find that these demands are so
restrictive that they constrain the MSSM parameter space in a way comparable to (though orthogonal to)
the typical assumptions that lead to scenarios like the CMSSM or mSUGRA, allowing a scan over allowed
parameter space to be feasible.

A large annihilation cross section to monoenergetic photons relative to that for continuum photons can be obtained within the MSSM if the LSP is a purely bino-like neutralino. In this case, the dominant cross section to continuum photons comes from $\chi \chi \rightarrow \bar{f}f$, a process that is always suppressed in some way.  As we demonstrate here, the suppression of the continuum cross section can be large enough that it becomes comparable to the cross section to monoenergetic photons, a scenario at odds with the intuition that tree-level annihilation to continuum photons will have a much larger cross section than loop-suppressed annihilation to final states with monoenergetic photons.

Although this type of analysis is relevant to an understanding of the implications of the Fermi-LAT data, its utility
extends beyond that particular observed excess,
which may simply be a statistical or systematic effect~\cite{Systematics}.
Rather, one can consider the implications of the detection of a line
signal with future experiments, such as GAMMA-400~\cite{gamma400} or \c{C}erenkov Telescope Array (CTA)~\cite{CTA}, or the emergence of such a
signal in the data from the Fermi satellite or ground-based gamma-ray detectors such as VERITAS~\cite{veritas} and HESS~\cite{hess}.
The ground-based CTA will have an effective area several orders of magnitude larger than the effective area of space-based telescopes,
but will suffer from the background of cosmic ray-induced \c{C}erenkov showers.
The GAMMA-400 satellite, by contrast is not expected to have a significantly larger effective area than the Fermi-LAT, but it
may have much better energy resolution.
Here we will assume that relevant experiments can distinguish a true line signal from a very hard continuum spectrum, as
might arise from a three-body annihilation process such as $\chi \chi \rightarrow f \bar f \gamma$.

In section~\ref{sec:genprinc}, we describe the general principles that lead to the region of MSSM parameter space of interest,
where the annihilation cross section to monoenergetic photon final states is comparable to that for final states that give
rise to a continuum spectrum of photons.
In section~\ref{sec:features} we analyze how the
cross section for annihilation to various final states scales with relevant model parameters.  In section~\ref{sec:scan} we
discuss a detailed numerical scan of the annihilation cross sections in this region of parameter space and relevance of our
results for more general non-MSSM scenarios.
Finally, we state our conclusions in section~\ref{sec:conclusions}.

\section{General Principles}
\label{sec:genprinc}

As put forward in~\cite{Cohen:2012me},
one can parameterize the
limits on dark matter annihilation obtained from gamma-ray observations in terms of
two quantities,
\bea
\sigma_{line} &=& 2\sigma_{\gamma \gamma} + \sigma_{\gamma Z},
\nonumber\\
R_{th} &=& {\sigma_{ann.} \over \sigma_{line}} ,
\eea
where $\sigma_{ann.}$ is the total dark matter annihilation cross section
and $\sigma_{(\gamma \gamma, \gamma Z)}$ are the $\chi \chi \rightarrow (\gamma \gamma, \gamma Z)$ annihilation
cross sections, respectively.  Since the neutralino LSP is a Majorana fermion, annihilation to the
$\gamma h$ final state is $p$-wave suppressed, and can therefore be ignored.
The rate at which dark matter annihilates to
monoenergetic photons is thus proportional to $\sigma_{line}$, and $R_{th}$ is the ratio of the total
annihilation rate to the rate of annihilation to monoenergetic photons.
If an excess of monoenergetic photons is observed, then, with some assumptions
about the dark matter density profile of the source, one can estimate the quantity $(\sigma_{line}v)$.
Assuming
the spectra of continuum photons from relevant dark matter annihilation processes are roughly similar in shape,
gamma-ray observations can then place an approximate upper bound on $R_{th}$.

Without any assumptions about the relationships between MSSM parameters, the number of free parameters is so vast
that a complete scan is impractical.  This consideration usually leads to assumptions
motivated by, among other things, naturalness, a string theoretic origin, a putative SUSY breaking mechanism, elegance or simplicity.
We will
instead consider a framework where the relations assumed between parameters are motivated only by the goal
of maximizing $(\sigma_{line}v)$ and minimizing
the annihilation to continuum photons.
We will find that these considerations are sufficient
to drastically reduce the parameter space of the MSSM.

We are guided by general considerations underlying the suppression of
annihilation to continuum photons
that is often
seen in MSSM models.
\begin{itemize}
\item{Since $\chi$ is a Majorana fermion, the cross section for annihilation to Standard Model
fermions, $\chi \chi \rightarrow \bar f f$ is suppressed either by $v^2 \sim 10^{-6}$ (for $p$-wave
annihilation) or by $m_f^2$ or the sfermion-mixing angle (for $s$-wave annihilation).  The
cross section for annihilation to the light Higgs, $\chi \chi \rightarrow hh$ is also $p$-wave
suppressed.}
\item{In the case of annihilation to fermions, the three-body annihilation processes
$\chi \chi \rightarrow \bar f f (\gamma, Z, g)$ can be important~\cite{EWbrem}.  These processes are typically suppressed
by a factor of $\alpha/r^2$ or $\alpha_s /r^2$, where $r= m_{\tilde f}^2 / m_\chi^2$ and
$\tilde f$ is the exchanged sfermion.}
\item{There is no chirality or $p$-wave suppression for $\chi \chi \rightarrow W^+ W^-$, $ZZ$, or $Zh$.
But these processes are only allowed at tree-level if $\chi$ has some wino or higgsino
component; if $\chi$ is purely bino, these processes do not occur at tree-level.}
\item{If $m_\chi > m_t$, then the annihilation process $\chi \chi \rightarrow \bar t t$ is kinematically
allowed.  There is no significant suppression of the $s$-wave annihilation process unless
$m_\chi \gg m_t$.}
\end{itemize}
Based on these facts, the low-energy particle content of interest for minimizing $R_{th}$ in the limit where tree-level
annihilations to gauge bosons are negligible
must include a purely bino-like LSP, negligible sfermion mixing, heavy sbottoms and scharms, and heavy neutral Higgsinos.
Although the production of on-shell tops is kinematically forbidden if $m_\chi < m_t$,
the production of off-shell tops can still provide a large contribution to $(\sigma_{ann.}v)$.
Similarly, if $m_\chi \gg m_t$, then although the the $\chi \chi \rightarrow \bar t t$ annihilation
process will be kinematically suppressed, the sensitivity of gamma-ray observations to $(\sigma_{line}v)$
will also be suppressed.
Finally, we note that if up/down/strange-squarks are light then the $\chi \chi \rightarrow \bar u u g$, $\bar d d g$,
and $\bar s s g$ three-body annihilation processes are only suppressed by $\alpha_s$.  To minimize $R_{th}$, we will
thus assume that all squarks are heavy.

We are therefore led to consider the scenario in which the only sfermions that are possibly light are
the sleptons.  The charged slepton, sneutrino, and LSP masses
are the only parameters that determine both $\sigma_{line}$ and $R_{th}$.
This scenario is consistent with constraints from the Large Hadron Collider, which has ruled out
models with light first generation squarks~\cite{SquarkSearch}.

\section{Features of This Scenario}
\label{sec:features}

\begin{figure*}[t]
\begin{center}
\vspace*{.2in}
\centering
\includegraphics[width=\textwidth]{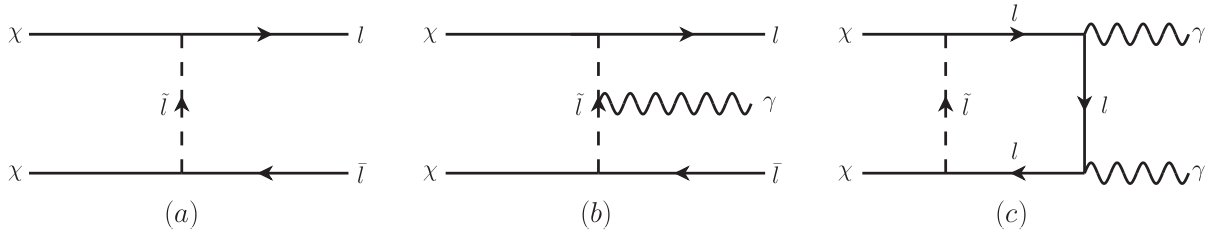}
\caption{Typical Feynman diagrams for the annihilation processes $\chi \chi \rightarrow \bar l l$ (left),
$\chi \chi \rightarrow \bar l l \gamma$ (center) and $\chi \chi \rightarrow \gamma \gamma$ (right).}
\label{fig:diagrams}
\end{center}
\end{figure*}

In this section we discuss the general model features
that determine the relevant quantities $R_{th}$ and $(\sigma_{line}v)$.
Since the LSP is largely bino-like and the squarks and Higgsinos are very heavy,
tree-level annihilation occurs largely through $t$-channel exchange of light sleptons
and sneutrinos, as shown in Fig.~\ref{fig:diagrams}(a).
The primary  two-body annihilation processes that contribute to $(\sigma_{ann.}v)$ in this case are
$\chi \chi \rightarrow \bar f f$ where $f=e,\mu ,\tau , \nu_{e, \mu ,\tau}$.
We parameterize the scale of the various annihilation processes in terms
of $\alpha$, $m_\chi$, and $r$, as defined above.
The cross sections
for these processes scale as
\begin{equation}
 \propto \left( \frac{\alpha^2}{m_{\chi}^2 \, r^2}\right) \times\left[ v^2 {\rm \,or\, } \left(\frac{m_f^2}{m_\chi^2}\right)\right].
 \end{equation}
Since $v^2 \approx 10^{-6}$ today, and $m_f^2 \ll m_\chi^2$, these processes are typically negligible.

The relevant three-body annihilation processes are $\chi \chi \rightarrow \bar l l (\gamma, Z)$,
$\bar \nu_l \nu_l Z$, $\bar l \nu_l W^-$, and $\bar \nu_l l W^+$, where $l=e, \mu ,\tau$.
These processes occur through $t$-channel exchange of a light sfermion, with
an electroweak gauge boson emitted either from the outgoing fermions or from the internal sfermion
line, as shown in Fig.~\ref{fig:diagrams}(b).  If the gauge boson is emitted from the internal line, then the diagram has two scalar propagators
and scales as $r^{-2}$.  There are two diagrams in which the gauge boson is emitted from the outgoing
fermions, and each diagram has only one scalar propagator.  But the diagrams partially cancel, and their
sum also scales as $r^{-2}$~\cite{Ciafaloni:2011sa}.
The electroweak bremsstrahlung processes are further suppressed by an additional factor of $\alpha$.
For $m_{\chi} \approx \frac{1}{2} m_{Z,W}$, the cross section
for annihilation to a three-body state with a massive vector boson will also be phase-space suppressed.
Finally, the cross section for annihilation to $\bar l l \gamma$ also receives an additional suppression due to
angular momentum conservation; for certain regions of the three-body final state phase space, the angular
momentum of the final state can only be zero if the gauge boson has helicity zero, which is not possible for a
massless gauge boson.  If $m_{\chi} > \frac{1}{2}m_{Z,W}$, then the total three-body annihilation
cross section will scale as
\begin{equation}
\propto \left(\frac{\alpha^2}{m_{\chi}^2 \, r^2}\right) \times \bigg[ \frac{\alpha}{r^2} \bigg].
\label{eq:3body}
\end{equation}

The processes $\chi \chi \rightarrow \gamma \gamma$, $\gamma Z$ proceed through box diagrams with $l,\tilde l$
running in the loop, as shown in Fig.~\ref{fig:diagrams}(c).  The number of slepton propagators in the loop can vary between
one and three, but as the sleptons become heavy, the dominant diagram will have only one slepton in the loop.
We thus find
\begin{equation}
(\sigma_{line}v) \propto \left(\frac{\alpha^2}{m_{\chi}^2 \, r^2}\right)\times \left[\alpha^2\right].
\end{equation}

The three-body annihilation process can dominate over $\chi \chi \rightarrow \bar l l$ only if
$\alpha / r^2 > v^2$.  Similarly, three-body annihilation can only dominate over $\chi \chi \rightarrow \bar b b$ if
$\alpha / r^2 > m_b^2 m_\chi^2 r^2/ m_{\tilde b}^4$.  Assuming that the squarks are very heavy, three-body
annihilation is the dominant continuum process if $r < {\cal O}(10^2)$.
As sleptons become heavier, however, annihilations to three-body final states become insignificant, and the two-body final states determine the annihilation cross section to continuum photons, $(\sigma_{cont.}v)$.  So we find

\begin{equation}
R_{th}-1 = \frac{(\sigma_{cont.}v)}{(\sigma_{line}v)} \sim
\begin{cases}
 1/(\alpha r^2) \sim {\cal O}(10^{-2}-10^2) &\text{ for } r \lesssim {\cal O}(10^2) \\
 v^2/\alpha^2 \sim  {\cal O}(10^{-2}) &\text{ for } r \gtrsim {\cal O}(10^2).
\end{cases}
\end{equation}
We thus expect to find values of $R_{th}$ ranging from $\sim 1$ to $\sim 100$.  But small $R_{th}$ arises
only when the sleptons are heavy, implying that the magnitude of $(\sigma_{line}v)$ will be suppressed.
Maximizing the magnitude of the line signal would require $R_{th} \gsim {\cal O}(100)$.

\section{Parameter Space Scan and Analysis}
\label{sec:scan}

We calculate the spectrum of supersymmetric particles, Higgs bosons, and dark matter observables
using SuSpect~\cite{Djouadi:2002ze}, FeynHiggs~\cite{Hahn:2010te}, and MicrOMEGAs~\cite{Belanger:2010gh}, with supplementary
calculations of the annihilation cross sections to three-body final states following~\cite{EWbrem,Fukushima:2012sp}.  Annihilations to 3-body final states will be most significant when sfermion mixing is negligible, so we compute this only for the first two generations.
We focus on consistent models that fall at least approximately into the scenario described
in section~\ref{sec:genprinc}, and consider the heavy scalars to be degenerate, with masses $m_0$ in the
range ($m_\chi+3$ TeV, 8 TeV).  We fix the Higgs mixing parameter, $\mu$, to be $\mu=m_0$, to ensure that the lightest
neutralino is nearly pure bino, and allow masses for the left- and right-handed\footnote{By ``left- and right-handed''
scalars, we mean the partners of the left- and right-handed Standard Model leptons.} $\tilde{e}$, $\tilde{\mu}$,
and $\tilde \nu_{e,\mu}$ in the range ($m_\chi$, $m_0$), with $\tilde{\tau}$ and $\tilde \nu_{\tau}$ heavier by at least 500 GeV such that 3-body annihilations involving the third generation are subdominant to those involving the first two generations.
Finally, the trilinear couplings, $A$, may take values in the range ($-5 m_0$, $5m_0$), with
larger $\left| A \right|$ preferred by the CMS~\cite{CMShiggs} and ATLAS~\cite{ATLAShiggs} measurements of $m_h \approx 126$ GeV.
For the parameter points we consider, we require the mass of the lighter CP-even Higgs boson to lie within the range
$123 {\rm ~GeV} < m_h < 129{\rm ~GeV}$.

\begin{figure*}[t]
\centering
\includegraphics[width=0.48\textwidth]{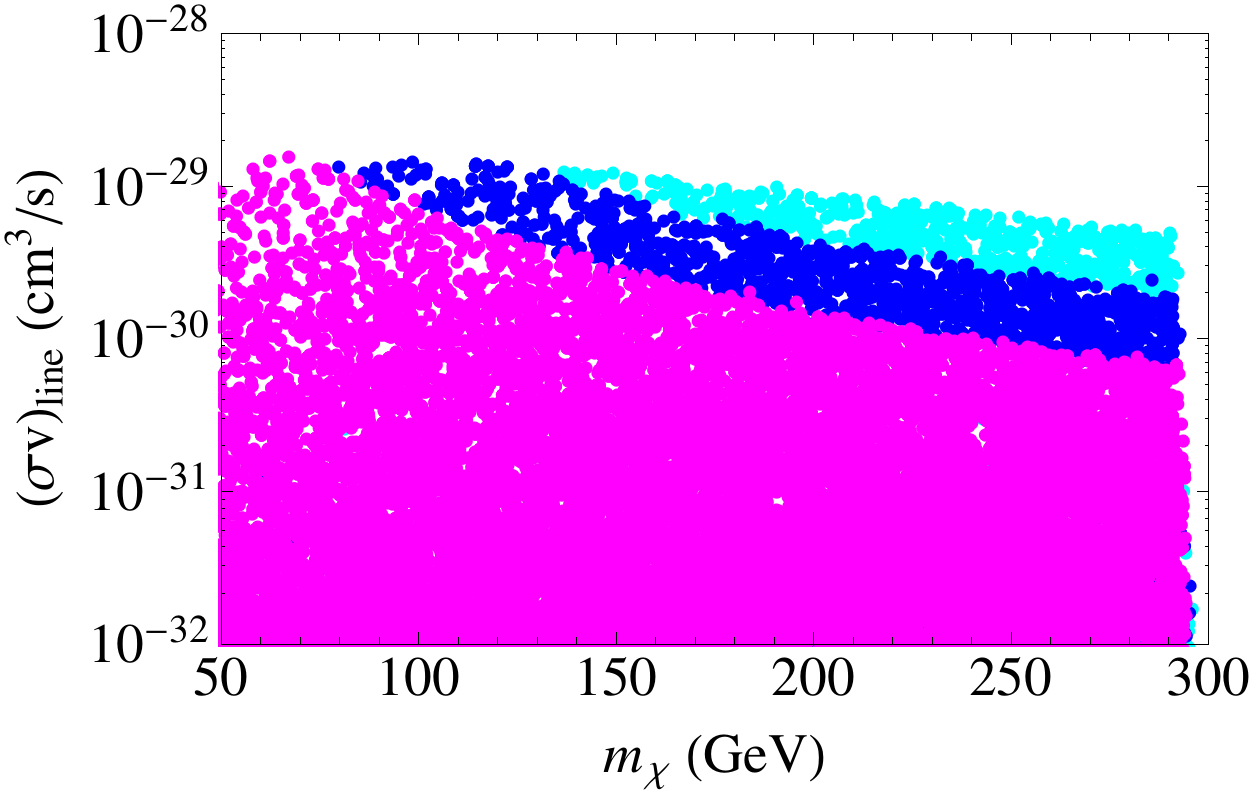}
\includegraphics[width=0.51\textwidth]{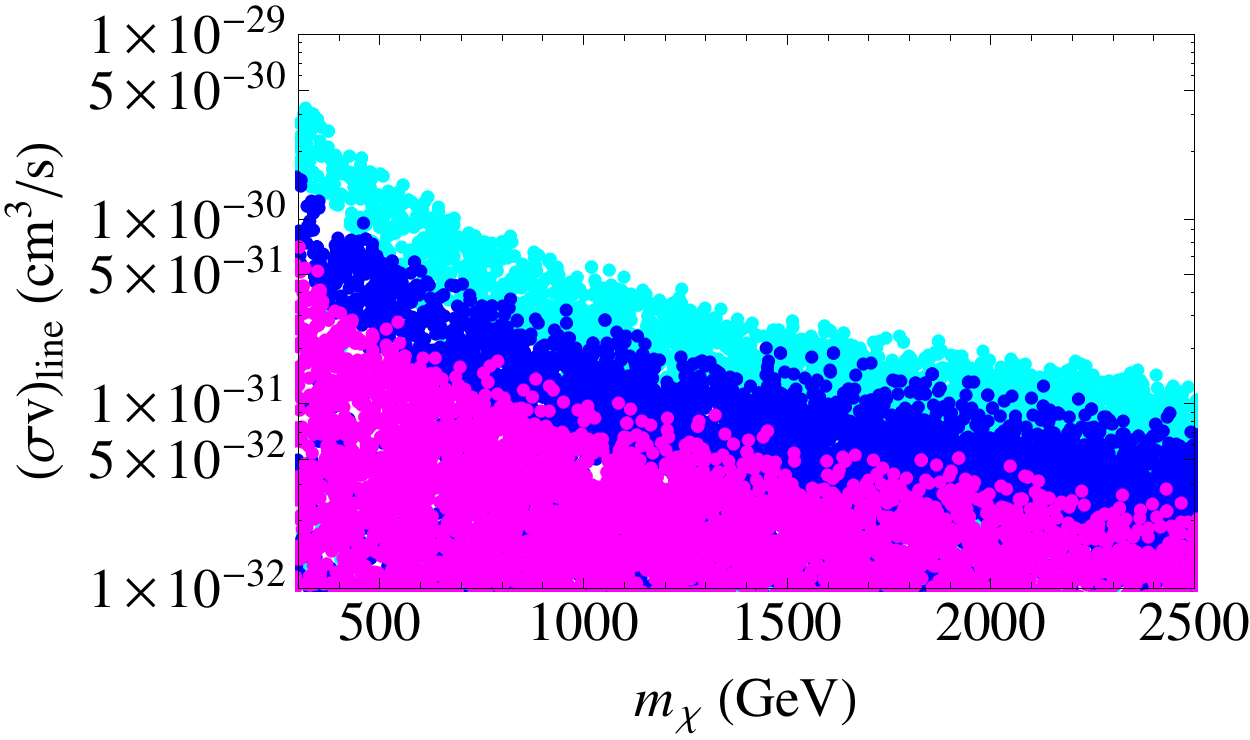}
\caption{Cross section for production of a line signal as a function of dark matter mass, colored by the value of $R_{th}$: $R_{th}<5$
(magenta), $5\leq R_{th}<10$ (blue), and $R_{th}\geq10$ (cyan).}
\label{fig:sigvline}
\end{figure*}

We assume that the lightest neutralino constitutes all of the dark matter in the universe.  In figure~\ref{fig:sigvline}, we plot the
cross section for production of a line signal, $(\sigma_{line} v)$, as a function of dark
matter mass.  The points are color-coded by the value of $R_{th}$: $R_{th}<5$ (magenta), $5\leq R_{th}<10$ (blue), and $R_{th}\geq10$ (cyan).  In the left panel, we address potential line signals observable by the Fermi satellite, for neutralino
masses up to roughly 300 GeV, while in the right panel we consider neutralino masses as large as 2.5 TeV.  Though HESS and VERITAS are sensitive
to dark matter masses in this range, the cross sections to monoenergetic photons are all well below the current limits~\cite{hess}.
We note that with adequate energy resolution, the $\gamma \gamma$ and $\gamma Z$ line signals will be distinguishable, and $R_{th}$ and
$(\sigma_{line}v)$ will not be optimal parametrizations of the signal strength.  Nonetheless, any observed line signal(s) can be simply mapped to particular values of $R_{th}$ and $(\sigma_{line}v)$.  

From the analysis of the previous section, we expect that in the decoupling limit discussed in the previous section, models with small $(\sigma_{line}v)$ have large values of $r$, which in turn implies small $R_{th}$.  This expectation is borne out in figure~\ref{fig:sigvline}: For $m_{\chi} \gtrsim 80$ GeV, smaller $R_{th}$ is possible at smaller $(\sigma_{line}v)$.  Furthermore, the smallest line cross sections (below $10^{-31}$ cm$^3$s$^{-1}$) do indeed come from points for which all sleptons are heavy.  However, $R_{th}$ is not necessarily small for these points unless the LSP is much heavier than the $\tau$ lepton, i.e.~$m_\tau^2/m_\chi^2\lesssim v^2$.  Such large $R_{th}$ points do not satisfy the limit discussed in the previous section.

Note also that for very light LSP's ($m_\chi \lesssim m_Z$), $R_{th}$ may be small even when $(\sigma_{line}v)$ is large.  In these cases, although annihilation to 3-body final states is the dominant mechanism for producing continuum photons, annihilation to 3-body final states with massive vector bosons is kinematically suppressed relative to the behavior approximated by Equation~\ref{eq:3body}. 

Finally, these largest cross sections, $(\sigma_{line}v) \approx 10^{-29}\text{ cm}^3/\text{s}$, are indeed obtained in the limit $m_{\tilde{l}} \approx m_\chi$, the limit discussed in Ref.~\cite{buckleyhooper}.
GAMMA-400 may be sensitive to cross sections of this size for small enough $m_\chi$~\cite{Bergstrom:2012vd}.
It has been shown that for $m_{\tilde{l}} \approx m_\chi$, very hard bremsstrahlung photons can mimic a gamma-ray line signal for the present
Fermi-LAT energy resolution~\cite{shakya}.
The analysis of the previous section shows that for $r \lesssim 10^2$, the ratio of the annihilation cross section to 3-body final states, $\sigma_{\text{3-body}}$, to that to 2-body final states that contribute to the continuum photon flux, $\sigma_{\text{2-body}}$, is 
\begin{equation}
\frac{\sigma_{\text{3-body}}}{\sigma_{\text{2-body}}} 
\sim \frac{\alpha}{r^2v^2}
\sim \frac{\alpha^2}{ v^2} (R_{th}-1) \sim 10^2 (R_{th}-1).
\end{equation}
  One expects that this ratio
can thus be as large as $\sim10^4$.  In this limit where annihilations to 3-body final states are most significant, i.e.~large $R_{th}$, we expect $\sigma_{\text{2-body}}$ : $\sigma_{line}$ : $\sigma_{\text{3-body}}$ to be $1$ : $10^2$ : $10^4$. 
In this analysis, all electroweak bremsstrahlung cross sections are included as part of the annihilation cross section to continuum photons, despite the fact that the spectrum may be significantly different from that for annihilations to 2-body final states.
Therefore one can interpret $R_{th}$ for the points with the largest cross sections in figure~\ref{fig:sigvline} as maximal,
since any very hard bremsstrahlung will effectively contribute to a line signal, as perceived by Fermi-LAT, and not to the perceived continuum. The ability to distinguish the 3-body continuum spectrum, the 2-body continuum spectrum, and a line signal may indeed be very important to determining the nature of particle dark matter.

\begin{figure*}[t]
\centering
\includegraphics[width=0.52\textwidth]{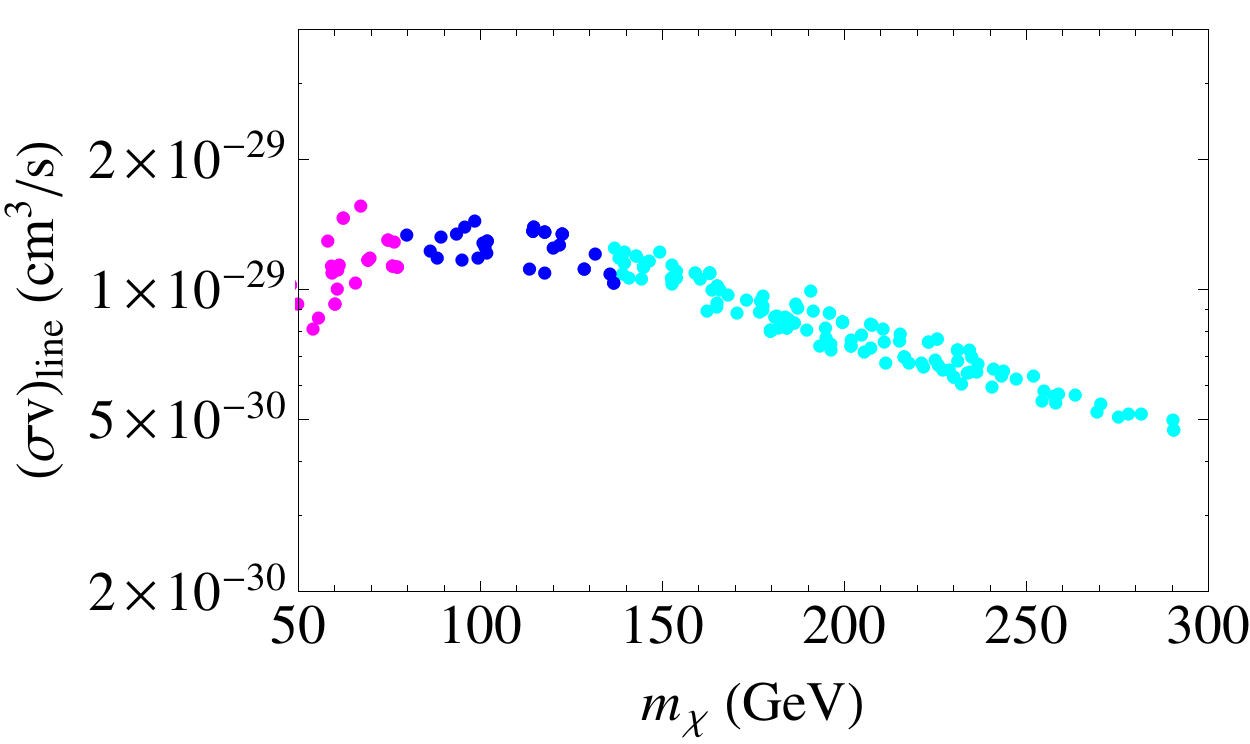}
\includegraphics[width=0.47\textwidth]{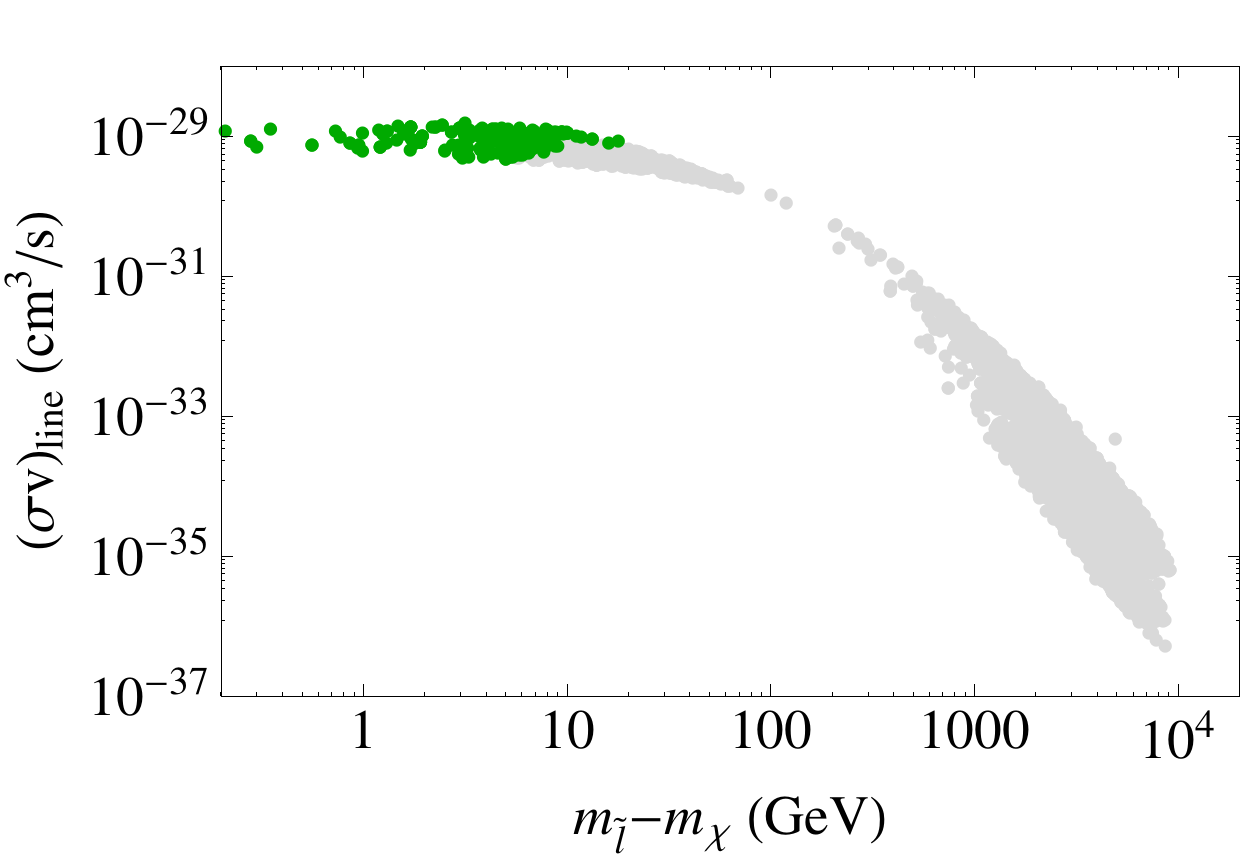}
\caption{The cross section for production of a line signal as a function of dark matter mass (left) and the mass difference between the lightest slepton and the neutralino LSP (right).  In the left panel, all points satisfy $\Omega_\chi h^2 \leq 0.15$, and are colored by the value of $R_{th}$: $R_{th}<5$ (magenta), $5\leq R_{th}<10$ (blue), and $R_{th}\geq10$ (cyan).  In the right panel, green points satisfy $\Omega_\chi h^2 \leq 0.15$, while light gray points predict $\Omega_\chi h^2 > 0.15$.}
\label{fig:sigvlineOmega}
\end{figure*}

In the left panel of figure~\ref{fig:sigvlineOmega} we focus on neutralinos with $m_\chi \leq 300$ GeV and plot only points for which the thermal relic abundance of dark matter satisfies
$\Omega_\chi h^2 \leq 0.15$, in rough compatibility with
the observed abundance of cold dark matter~\cite{wmap}.
We see that some of the very largest line cross sections are due to models in which the dark matter is a thermal relic.  It is not surprising that we find points for which $R_{th} \sim {\cal O} (10^2)$ with $( \sigma _{line}v) \sim 10^{-29}$ cm$^3/$s, i.e.~a
continuum cross section today of $(\sigma_{cont.}v) \sim 10^{-27}$ cm$^3/$s, and possibly even a larger annihilation cross section in the early universe, allowing compatibility between the measured dark matter abundance and the thermal relic density. Indeed, there are even points for which $R_{th}<5$ that can predict a relic abundance of neutralinos compatible with
the observed dark matter abundance.   In fact, these models are precisely those in which the annihilation rate in the early universe is significantly enhanced due to coannihilations of neutralinos with light sleptons.  This is evident in the right panel of figure~\ref{fig:sigvlineOmega}, where we plot 
$(\sigma_{line} v)$ as a function of the mass splitting between the neutralino LSP and the lightest slepton.  Green points satisfy $\Omega_\chi h^2 \leq 0.15$, while light gray points predict $\Omega_\chi h^2 > 0.15$.  For $m_\chi \lesssim 80$ GeV, there are models with small $R_{th}$ and $(\sigma_{line} v) \gtrsim 10^{-29}$ cm$^3/$s that provide the best prospects for detection.


Although this analysis has been presented in the framework of the MSSM, it applies more generally
to any model where the dark matter candidate is a Majorana fermion that is neutral under the
Standard Model gauge groups.  For such a model, dark matter can annihilate at tree-level only
to $\bar f f$ or $hh$, with the latter process being either $p$-wave suppressed or suppressed by CP-violating
phases.  If annihilation to $\bar f f$
proceeds through $t$-channel exchange of a scalar partner, then the analysis given here for the
relation between $R_{th}$ and the mass of the scalar partner still holds.  Note, however, that
the bino-fermion-sfermion coupling is simply related to the
hypercharge coupling.  For a more general dark matter model, this coupling, $g$, is a free parameter.
Since $(\sigma_{line}v)$ and all continuum annihilation cross sections scale as $g^4$, the choice of
coupling does not affect $R_{th}$.  However, it does affect the overall scale of the
annihilation cross section; if this coupling is required to be perturbative, then
$\sigma_{(\gamma \gamma,\gamma Z)}$ can be increased by up to 4 orders of magnitude, in agreement with the limit
discussed in Ref.~\cite{buckleyhooper}.  For models
with small $R_{th}$, such an enhancement could bring a line signal within the reach of GAMMA-400.

\section{Conclusions}
\label{sec:conclusions}

We have considered a class of MSSM models for which the ratio $\sigma(\chi \chi \rightarrow \gamma \gamma, \gamma Z) /
\sigma(\chi \chi \rightarrow {\rm anything})$ is maximized.  We have found that this requirement leads to
a dramatic restriction in the number of relevant MSSM parameters.
In particular, the region of interest studied here contains a nearly pure bino LSP, with the only light superpartners being the bino
and the sleptons.
Interestingly, this region of parameter space is consistent with recent bounds from
the LHC, which tightly constrain models with light squarks.

For such models, we find that if a photon line signal is potentially observable at current or next generation experiments,
then $R_{th}$ can be as low as ${\cal O}(10-100)$.  Moreover, for many such models
the continuum signal is dominated by three-body annihilation.  Since three-body annihilation may produce a very hard
photon spectrum, it is possible that such a continuum annihilation process could be mistaken for a line.  Therefore our
numerical results for $R_{th}$ represent the maximal values possible (i.e.~the limit of perfect energy resolution).
The improved energy resolution of next generation gamma-ray telescopes should make it possible to resolve this difference.

For MSSM models with small $R_{th}$, the line signal will be a few orders of magnitude too small to be
observed with Fermi.  However, more general models of Standard Model-neutral Majorana fermion dark matter can achieve
the same $R_{th}$ with an enhancement in the line signal of up to four orders of magnitude, potentially
bringing a pure line signal with no observable continuum spectrum within reach of observation.

\vskip .2in
\textbf{Acknowledgments}

We thank the organizers of ICHEP2012 and the Center for Theoretical Underground Physics and
Related Areas (CETUP* 2012) in South Dakota
for their support and hospitality during the completion of this work.  We are grateful to
D.~Marfatia and C.~Kelso for useful discussions, and to K.~Fukushima for comments and
for assistance with the numerical code for computing the continuum annihilation cross section. 
We are also grateful to the anonymous referee for insightful suggestions.
Support and resources from the Center for High Performance Computing at the University of Utah are also gratefully acknowledged.
The work of J.~K.~is supported
in part by Department of Energy grant DE-FG02-04ER41291.

\end{document}